\documentclass[sigconf]{acmart}

\usepackage{booktabs}
\usepackage{multirow}
\usepackage{graphicx}

\setcopyright{none}
\acmConference[SeT-LLM @ KDD 2026]{2nd Workshop on Secure and Trustworthy Large Language Models}{August 9--10, 2026}{Jeju, South Korea}
\acmBooktitle{2nd Workshop on Secure and Trustworthy Large Language Models (SeT-LLM @ KDD 2026)}
\renewcommand\footnotetextcopyrightpermission[1]{}
\newcommand{\cascadeDeferralPct}{89\%}

\begin{document}

\title{A Four-Axis Trustworthiness Benchmark\\for LLM-as-Judge in Principle-Based Regulation}

\author{Dipankar Sarkar}
\email{dipankar@skelfresearch.com}
\affiliation{%
  \institution{Independent Researcher}
  \country{}
}

\begin{abstract}
\emph{Principle-based} regulation, with evaluative standards such as ``fair,
clear, and not misleading'' or ``deliver good outcomes'', cannot be reduced to
binary predicates, and LLM-as-judge is increasingly used as the substitute. Our
position is that any such judge must be evaluated on four axes:
\emph{accuracy}, \emph{paraphrase robustness}, \emph{adversarial robustness},
and \emph{calibration}. We release \textsc{Principle-Bench}, 168 cryptoasset
financial-promotion scenarios mapped to two UK FCA principles, with paraphrase,
adversarial keyword-stuffing, and boundary perturbations authored under a
pre-registered rubric; the first benchmark covering all four axes for
principle-based regulation. We also introduce \textsc{Ceca} (Calibrated
Exemplar-Cluster Assessment): a calibrated, auditable assessor that emits exact
per-exemplar counterfactual attributions. Across keyword counting, three
sentence-transformer embedders, an open-weight LLM-judge, and a calibrated
cascade, no method dominates all four axes. A 120B LLM-judge, strongest on
benign inputs, loses 47 accuracy points (0.74\,$\to$\,0.27) on keyword-stuffed
Consumer Duty inputs: ``compliance theatre.'' A second judge from a different
model family agrees only at Cohen's $\kappa=0.16$ on that split, localising the
failure to the model rather than the corpus. Any deployment-grade LLM-judge for
principle-based regulation must report per-principle adversarial deception and
post-hoc calibration alongside aggregate accuracy.
\end{abstract}


\maketitle

\section{Introduction}

Automated compliance checking is now a staple of AI governance pipelines in
finance, healthcare, and autonomous systems. The dominant paradigm treats every
regulation as a Horn clause: a binary predicate that is either satisfied or
violated. This rule-based model underpins Datalog verifiers, semantic-graph
checkers, and most production reg-tech stacks.

However, not every regulatory measure is a rule. The UK Financial Conduct
Authority's Consumer Duty and COBS~4.5A.3R stipulate that promotions be ``fair,
clear, and not misleading'' and that firms ``deliver good outcomes for retail
customers'' \citep{fca2022consumerduty}. These are \emph{principles}:
outcome-based, evaluative standards that cannot be verified by checking a single
boolean. A promotion can contain every required disclaimer and still be
misleading overall. When rule-based systems meet principles, they either
silently treat them as rules (false confidence) or ignore them (vacuous
compliance). Both are dangerous in high-stakes domains. The EU AI Act
\cite{eu2024aiact} classifies credit-scoring and insurance-pricing as
high-risk, yet its conformity framework relies on prescriptive checks that do
not capture ``fairness'' or ``adequate human oversight.''

An obvious alternative is to ask a large language model to judge the principle
directly. LLM-as-judge \cite{zheng2023llmjudge} reaches near-human agreement on
benchmark tasks, but remains opaque, is poorly calibrated by default, and offers
no per-example justification a regulator could interrogate. The question for
secure and trustworthy deployment is not whether an LLM \emph{can} evaluate a
principle, but whether its evaluations are auditable, reproducible, calibrated,
and robust to gaming. This is the four-axis question the rest of the paper
makes operational.

\section{Contributions}

\textbf{(1)}~\textsc{Principle-Bench}, the first benchmark with a perturbation
suite built specifically for principle-based regulatory assessment: 168
scenarios across two UK FCA principles (100 original + 30 paraphrase + 28
adversarial + 10 boundary), anchored on author-prepared summaries of public FCA
material and labelled under a \emph{pre-registered} rubric whose git tag
pre-dates all labelling.
\textbf{(2)}~\textsc{Ceca}, a calibrated exemplar-cluster assessor with exact
per-exemplar counterfactual attribution: an auditable, transparency-by-design
alternative to an opaque judge (Appendix~\ref{app:method}).
\textbf{(3)}~A four-axis comparison of keyword counting, three
sentence-transformer embedders, an open-weight LLM-judge (with a second judge
from a different model family), and a calibrated cascade. \emph{No method
dominates.} Headline finding: a 120B LLM-judge drops 47 accuracy points
(Consumer Duty: 0.74\,$\to$\,0.27) on keyword-stuffed inputs, the
``compliance theatre'' legal scholarship warns of; a different-family judge
confirms the failure is model-side ($\kappa=0.16$).

\section{\textsc{Principle-Bench} and \textsc{Ceca}}

\paragraph{Problem.}
Let $x$ be input text and $P=(C,\tau)$ a principle with sub-concept clusters
$C=\{c_1,\dots,c_k\}$ and escalation threshold $\tau$. Each cluster
$c_i=(E^+_i,E^-_i,w_i)$ carries positive and negative exemplars and a weight. An
\emph{assessment} is $\alpha=(s,\{a_i\},e)$: a calibrated compliance probability
$s\in[0,1]$, per-cluster confidences $a_i$, and an escalation flag
$e=\mathbf{1}[s<\tau]$. It is \emph{auditable} if every component traces to a
specific exemplar contribution a regulator could remove or contest. We evaluate
assessors on four axes:
\textbf{accuracy}, \textbf{paraphrase robustness} (accuracy preserved under
exemplar-token-free rewrites), \textbf{adversarial robustness} (accuracy
preserved under surface keyword insertion designed to invert the verdict), and
\textbf{calibration} (low Expected Calibration Error, ECE).

\paragraph{\textsc{Principle-Bench}.}
168 scenarios across COBS~4.5A.3R (``fair, clear and not misleading'') and
PRIN~2A/FG22--5 Consumer Duty (``deliver good outcomes''), in four splits:
\emph{original} (100), \emph{paraphrase} (30, $\le5$ shared content tokens with
any exemplar, script-verified), \emph{adversarial} (28 keyword-stuffed; 2 of 30
refused by the generator's safety filter), and \emph{boundary} (10
near-threshold). Scenarios are generated by an open-weight model
(\texttt{gpt-oss:120b}) from one of 22 author-prepared summaries of public FCA
material, with a target verdict (the ground truth) and a ban on verbatim
exemplar phrases. The per-cluster rubrics were committed and tagged
(\texttt{v0-rubric-prereg}) \emph{before} any scenario was scored, giving a
tamper-evident pre-registration record. Construction, limitations (author-only
labelling and a corpus/judge model-family overlap, partly mitigated in
\S\ref{sec:results}), and the reading list are detailed in
Appendix~\ref{app:bench}.

\paragraph{\textsc{Ceca}.}
On a pluggable embedder $\phi$, the raw cluster confidence is a sigmoid of the
positive-minus-negative mean cosine similarity to exemplars, and the
principle-level raw score is their weighted mean (Appendix~\ref{app:method},
Eq.~\ref{eq:cluster}--\ref{eq:overall}); averaging makes the score Lipschitz in
any single exemplar, bounding non-adversarial sensitivity. Raw scores occupy a
narrow sub-interval of $[0,1]$, so we fit a logistic $s=\sigma(a\hat
s+b)$ (Platt scaling) on a 20-scenario dev split; the calibrated $s$ is what a
regulator sees. Because the assessor is a closed-form function of the exemplar
set, \textsc{Ceca} emits an \emph{exact} per-exemplar counterfactual
$\Delta_e=s-s_{\setminus e}$: which exemplar drove the verdict, and what removing
it would do. A \emph{cascade} defers to the LLM-judge only when the primary's
confidence lands in a dev-tuned uncertainty band, bounding judge invocations
while keeping attribution for the high-confidence majority.

\section{Experimental Setup}

We compare six methods: \textbf{Keyword} (deterministic bag-of-clusters, the
state of practice); three sentence-transformer embedders, \textbf{MiniLM}
(\texttt{all-MiniLM-L6-v2}), \textbf{BGE} (\texttt{bge-base-en-v1.5}), and
\textbf{Ollama-mxbai} (\texttt{mxbai-embed-large}); \textbf{LLM-judge}
(\texttt{gpt-oss:120b}, temperature~0, seed~42, prompted with the principle
text, full exemplar set, and input); and a \textbf{cascade} (keyword primary,
LLM-judge fallback in band $[0.4,0.6]$). Calibrators and the cascade band are
tuned on a 20-scenario dev subset of \emph{original}; all metrics use the
remaining 80 plus the perturbation splits. For 7--16\% of \emph{original}
scenarios the judge returns malformed JSON; those cells are computed on retained
scenarios (the $n$ column). Point estimates carry bootstrap 95\% CIs ($10^4$
resamples); paired comparisons use McNemar's exact test. Full method, metric,
and prompt details are in Appendix~\ref{app:setup}.

\section{Results}\label{sec:results}

\begin{figure}[t]
\centering
\includegraphics[width=0.95\columnwidth]{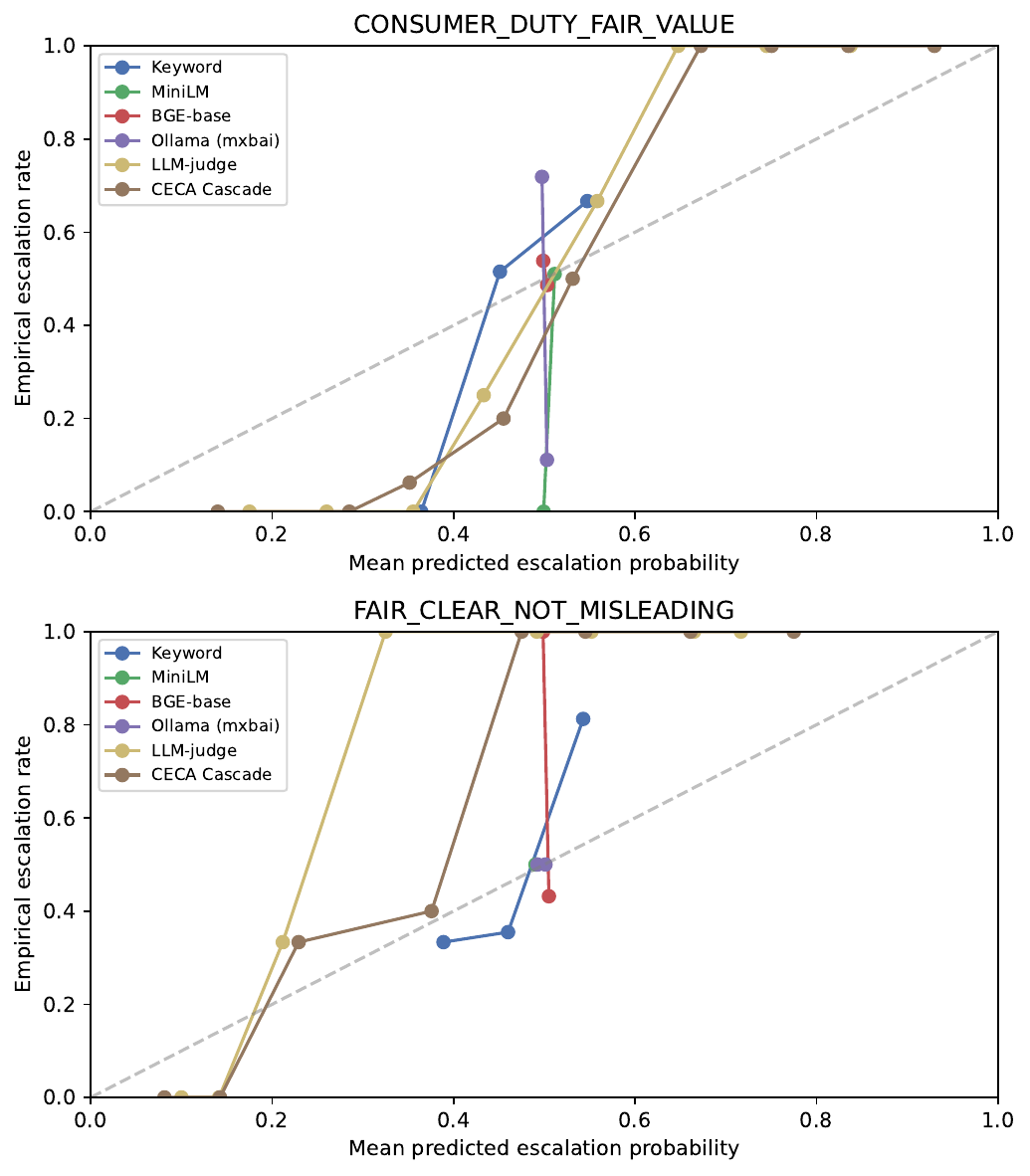}
\caption{Reliability diagrams on the \emph{original} split. Sentence-transformer
methods trace near-vertical curves (raw scores compress around 0.5); the
LLM-judge over-promises pre-calibration. Post-Platt scaling on a 20-sample dev
split, the LLM-judge's test reliability aligns with the diagonal.}
\label{fig:reliability}
\end{figure}

\paragraph{No method dominates (Table~\ref{tab:primary}).}
On \emph{original} COBS, LLM-judge and cascade are near-indistinguishable
(accuracy 0.96 vs.\ 0.94; AUC 1.00 vs.\ 0.99) and dominate the embedders, which
sit at chance because their raw scores compress around 0.5. On the harder
Consumer Duty principle the gap widens (LLM-judge 0.74, cascade 0.67). After
Platt scaling, the LLM-judge's deployment-relevant test ECE is 0.04--0.10,
down from a raw 0.22--0.27 (Fig.~\ref{fig:reliability}); calibration is not
optional. Paraphrase robustness favours the LLM-judge and cascade; the
embedders, already at chance, have little left to lose
(Appendix~\ref{app:results}, Fig.~\ref{fig:appfigs}).

\begin{table}[t]
\centering\small
\setlength{\tabcolsep}{4pt}
\begin{tabular}{llcccc}
\toprule
\textbf{Principle} & \textbf{Method} & \textbf{Acc.\,[95\% CI]} & \textbf{F1} & \textbf{AUC} & \textbf{ECE} \\
\midrule
\multirow{6}{*}{\rotatebox{90}{Cons.\ Duty}}
 & Keyword      & 0.54 [.40,.68] & 0.68 & 0.81 & 0.196 \\
 & MiniLM       & 0.50 [.36,.64] & 0.67 & 0.61 & 0.128 \\
 & BGE-base     & 0.50 [.36,.64] & 0.67 & 0.54 & 0.109 \\
 & Ollama-mxbai & 0.50 [.36,.64] & 0.67 & 0.17 & 0.106 \\
 & LLM-judge    & \textbf{0.74} [.62,.87] & 0.80 & 0.98 & 0.096 \\
 & Cascade      & 0.67 [.52,.79] & 0.75 & 0.96 & 0.113 \\
\midrule
\multirow{6}{*}{\rotatebox{90}{COBS 4.5A}}
 & Keyword      & 0.52 [.38,.66] & 0.67 & 0.70 & 0.108 \\
 & MiniLM       & 0.50 [.36,.64] & 0.67 & 0.42 & 0.329 \\
 & BGE-base     & 0.50 [.36,.64] & 0.67 & 0.18 & 0.132 \\
 & Ollama-mxbai & 0.50 [.36,.64] & 0.67 & 0.20 & 0.249 \\
 & LLM-judge    & \textbf{0.96} [.90,1.0] & 0.96 & 1.00 & 0.040 \\
 & Cascade      & 0.94 [.85,1.0] & 0.94 & 0.99 & 0.084 \\
\bottomrule
\end{tabular}
\caption{Primary metrics on the \emph{original} split (bootstrap 95\% CIs,
$10^4$ resamples). ECE is held-out test ECE after Platt scaling fit on a
20-scenario dev split. Embedders sit at chance; the LLM tier dominates on benign
inputs; a picture the adversarial split overturns.}
\label{tab:primary}
\end{table}

\begin{figure}[t]
\centering
\includegraphics[width=0.92\columnwidth]{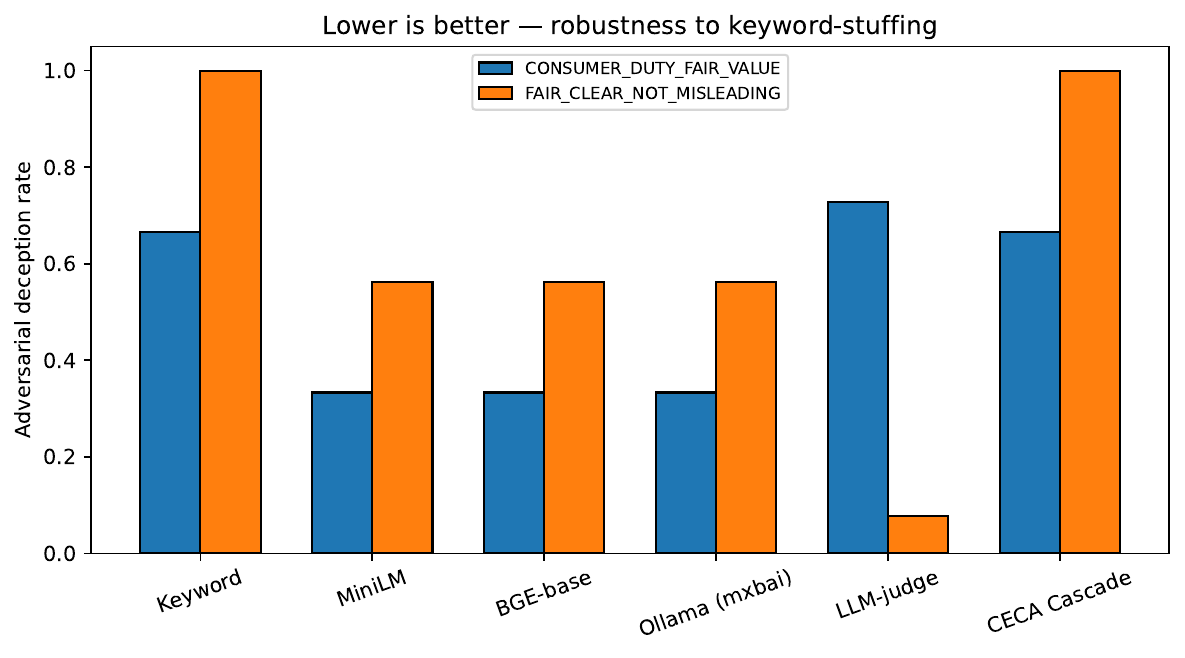}
\caption{Adversarial deception rate (verdicts flipped against ground truth on
keyword-stuffed inputs; lower is better). Consumer Duty \emph{inverts} the
benign ranking of Table~\ref{tab:primary}: the LLM-judge (the strongest method
on every other split) becomes the weakest (47-point accuracy drop),
while uncalibrated sentence-transformer embedders, blind to the injected
phrases, become the most robust.}
\label{fig:adversarial}
\end{figure}

\paragraph{The judge gets gamed (compliance theatre).}
The adversarial split inverts the ranking (Fig.~\ref{fig:adversarial}). The
LLM-judge (strongest everywhere else) collapses on Consumer Duty from 0.74 to
\textbf{0.27}, a
47-point drop, while holding 0.92 on COBS. The Consumer Duty perturbation
inserts short factual assertions (e.g.\ ``Client can absorb a total loss'')
verbatim into substantively non-compliant promotions; the judge over-credits
these surface statements and passes the promotion. COBS positive exemplars are
higher-order qualitative claims (``presents a fair picture'') that resist such
injection. This is the \emph{compliance theatre} legal scholarship flags
for principle-based regulation: game the assessor by stuffing the right phrases
while leaving practice unchanged. The uncalibrated sentence-transformer
embedders are the \emph{most} adversarially robust on Consumer Duty (0.67)
precisely because they cannot read the inserted phrases as evidence; keyword
fails 100\% on COBS adversarial by construction.

\paragraph{The failure is model-side, not a corpus artefact (Table~\ref{tab:kappa}).}
Because the corpus generator and one judge share a model family, we ran a second
judge from a \emph{different} family (\texttt{kimi-k2.6}) on all splits. On COBS,
inter-judge agreement stays high on \emph{original} ($\kappa=0.85$) and
\emph{paraphrase} (0.86); on Consumer Duty \emph{adversarial} it collapses to
$\kappa=0.16$, at chance. Two judges from different families produce
materially different verdicts on the same gamed inputs, so the susceptibility
to keyword-stuffing is model-side rather than an artefact of who authored the
corpus.

\begin{table}[t]
\centering\small
\begin{tabular}{lcc}
\toprule
\textbf{Split} & \textbf{COBS 4.5A.3R} & \textbf{Consumer Duty} \\
\midrule
\texttt{original}    & 0.853 & 0.493 \\
\texttt{paraphrase}  & 0.857 & 0.759 \\
\texttt{boundary}    & 0.600 & 1.000 \\
\texttt{adversarial} & 0.572 & \textbf{0.157} \\
\bottomrule
\end{tabular}
\caption{Inter-judge agreement (Cohen's $\kappa$) between \texttt{gpt-oss:120b}
and a different-family judge \texttt{kimi-k2.6}. The collapse to $\kappa=0.16$ on
Consumer Duty adversarial localises the keyword-stuffing vulnerability to the
model, not the corpus. Boundary $\kappa$ rests on $n=8$ and is a summary, not a
significance test.}
\label{tab:kappa}
\end{table}

\paragraph{Auditability and the cascade.}
\textsc{Ceca}'s counterfactual attribution is faithful: removing the
top-attributed exemplar shifts the score by $10\times$ a random exemplar's
effect and flips the verdict on 4\% [CI 1--8\%] of keyword-backend inputs;
enough to let a regulator ask ``which exemplar drove this, and what if it were
removed?'' The cascade (dev-tuned band, \cascadeDeferralPct{} deferral)
approaches LLM-judge accuracy on benign and paraphrased inputs (gaps 2--14
points) but \emph{inherits its primary's vulnerability}: on COBS adversarial it
scores 0\% because the keyword primary confidently mis-scores stuffed inputs
outside the band and never consults the judge. This is a structural hazard for
\emph{any} cascade whose primary is the direct target of the perturbation. Full
per-cluster AUC, cascade cost, calibration diagrams, and a worked attribution
example are in Appendix~\ref{app:results}.

\section{Discussion}

\paragraph{Why ``just use the LLM'' is unsafe.}
A naive reading of Table~\ref{tab:primary} says ``use the judge.'' Three
counter-observations: (i)~its raw confidences are polarised, inflating raw ECE to
0.22--0.27, so without Platt scaling a regulator cannot read the verdict's
confidence as a probability; (ii)~it offers narrative rationale, not the
ranked, exactly-reproducible per-exemplar contributions auditability requires,
so by our own criterion the judge is a strong \emph{baseline}, not a deployable
assessor; (iii)~it is gameable on at least one principle, with the failure
confirmed model-side.

\paragraph{Recommendation.}
Any deployment-grade LLM-judge for principle-based assessment should report a
per-principle \emph{adversarial deception rate} and \emph{post-hoc calibration}
alongside aggregate accuracy, and pair the verdict with an auditable,
contestable attribution layer. The method and benchmark structure
(pre-registered rubric, seed grounding, perturbation suite) transfer to any
principle-based domain: healthcare's ``informed consent,'' autonomous vehicles'
``reasonable safety,'' or the EU AI Act's evaluative standards, conditional on
a regulator-led exemplar-curation process. We release prompts,
raw responses, calibrators, and the pre-registration tags at
\url{https://github.com/sarkar-dipankar/principle-bench} (Appendix~\ref{app:repro}).

\section{Conclusion}

Principle-based regulation needs an assessor that is more than ``correct on
average'': calibrated, paraphrase- and adversary-robust, and counterfactually
auditable. Across all four axes no single method wins, and the strongest judge
is the one most easily gamed, a trustworthiness failure invisible to headline
accuracy. Secure-and-trustworthy LLM evaluation for governance must be grounded
in this multi-axis evidence.

\bibliographystyle{ACM-Reference-Format}
\bibliography{references}

\clearpage
\appendix
%% Appendix — unlimited length per SeT-LLM CFP (4pp main limit excludes refs + appendix).

\section{Related Work}\label{app:related}

\paragraph{Principles versus rules.}
Dworkin~\cite{dworkin1967model} argues rules operate all-or-nothing whereas
principles carry comparative weight; Black~\cite{black2008forms,black2010rise}
traces the rise and post-crisis retreat of principles-based financial
regulation, and Braithwaite~\cite{braithwaite2002rules} shows principles are
needed when rules over- or under-reach. The implication for automated
compliance is direct: a system limited to encoding rules cannot evaluate
principles, however advanced its rule engine.

\paragraph{Automated compliance and legal NLP.}
Sergot et al.~\cite{sergot1986british} represented the British Nationality Act
in Horn clauses; Governatori and Rotolo~\cite{governatori2008algorithm} added
defeasible reasoning; Athan et al.~\cite{athan2013oasis} formalised LegalRuleML;
Hashmi~\cite{hashmi2015methodology} keys on modal verbs (``must,'' ``shall'')
that principle-based regimes do not use. On the learning side, Reimers and
Gurevych~\cite{reimers2019sentencebert} introduced Sentence-BERT, Chalkidis et
al.~\cite{chalkidis2020legalbert} specialised BERT for legal text, and
\emph{LegalBench}~\cite{guha2023legalbench} curates legal-reasoning tasks for
LLMs; we depart from this work by targeting \emph{evaluative} regulation and a
purpose-built perturbation suite.

\paragraph{Prototypes, calibration, and LLM-as-judge.}
\textsc{Ceca}'s exemplar clusters extend prototypical
networks~\cite{snell2017prototypical,luo2023prototype} and prototype-based
explanation~\cite{chen2019protopnet} to regulatory text. Guo et
al.~\cite{guo2017calibration} show modern classifiers are overconfident and
post-hoc scaling helps; Zheng et al.~\cite{zheng2023llmjudge} find LLM judges
near human agreement on benchmark tasks but biased. We further find them
poorly calibrated and adversarially gameable in regulatory settings. Work on auditability and accountability (Mehrabi et
al.~\cite{mehrabi2021survey}, Arrieta et al.~\cite{arrieta2020explainable},
Mitchell et al.~\cite{mitchell2021algorithmic}, and the NIST~\cite{nist2022bias}
and ACM~\cite{acm2022principles} frameworks) motivates our
transparency-by-design per-exemplar attribution over post-hoc rationalisation.

\section{\textsc{Ceca} Method Details}\label{app:method}

\textsc{Ceca} is built in four layers on a pluggable embedder
$\phi:\mathit{String}\to\mathbb{R}^d$.

\paragraph{Base exemplar-cluster scoring.}
With $\cos(u,v)=u\cdot v/(\|u\|\,\|v\|)$ and $\sigma(z)=1/(1+e^{-z})$, the raw
cluster confidence is
\begin{equation}
\begin{split}
\rho(c,x)=\sigma\!\Big(\,&\textstyle\frac{1}{|E^+|}\sum_{e\in E^+}\cos(\phi(x),\phi(e))\\[-2pt]
                  -\;&\textstyle\frac{1}{|E^-|}\sum_{e\in E^-}\cos(\phi(x),\phi(e))\Big)
\end{split}
\label{eq:cluster}
\end{equation}
and the principle-level raw score is the weighted mean
\begin{equation}
\hat{s}(P,x)=\frac{\sum_{c\in C}w_c\,\rho(c,x)}{\sum_{c\in C}w_c}.
\label{eq:overall}
\end{equation}
Averaging (rather than maximising) makes $\rho$ robust to a single spurious
exemplar match. Assuming unit-normalised embeddings ($\|\phi(\cdot)\|=1$, the
standard sentence-transformer convention), $\rho$ is Lipschitz in any single
exemplar: replacing $\phi(e_j)\in E^+$ by $\phi(e_j)+\delta$ with
$\|\delta\|\le\epsilon$ shifts the positive-cosine average by at most
$\epsilon/|E^+|$, so $|\Delta\rho|\le\sigma'(\cdot)\cdot\epsilon/|E^+|\le
\epsilon/(4|E^+|)$ (using $\sigma'\le1/4$); the symmetric bound
$\epsilon/(4|E^-|)$ holds for negative exemplars, and $\hat{s}$ inherits the
bound through a convex combination. The empirical adversarial-deception rate is
the worst-case figure when the perturbation is a deliberate text-level
keyword-stuffing edit rather than an infinitesimal embedding shift.

\paragraph{Platt calibration.}
Raw scores are sigmoids of cosine differences and occupy a narrow sub-interval
of $[0,1]$. We fit a logistic $s=\sigma(a\hat{s}+b)$ on a held-out development
split (Newton--Raphson on cross-entropy; coefficients persisted to disk). The
calibrated $s$ is what a regulator sees; $\tau$ is then an empirical-probability
cut-off, not an arbitrary sigmoid output. We also report isotonic regression as
a robustness check.

\paragraph{Counterfactual attribution.}
For an assessment $(s,\{a_i\})$, \textsc{Ceca} emits, for each exemplar
$e\in E^+_i\cup E^-_i$, the counterfactual $s_{\setminus e}$ obtained if $e$ were
removed, and the signed delta $\Delta_e=s-s_{\setminus e}$. Because the assessor
is a closed-form function of the exemplar set, the predicted $s_{\setminus e}$
exactly matches re-running the assessor with $e$ removed. Faithfulness is
stronger than this identity: does the highest-$|\Delta|$ exemplar carry the
verdict? Empirically (\S\ref{app:results}), removing the top-attributed exemplar
flips the verdict on 4\% [95\% CI 1--8\%] of inputs and shifts the score by
$10\times$ the mean shift of a random exemplar.

\paragraph{Cascade with LLM-as-judge.}
Any embedder producing a per-cluster score in $[0,1]$ can be the primary. When
the primary's confidence $s$ falls in a dev-tuned band $[\ell,h]$, the cascade
defers to an LLM-judge with access to the same principle and exemplars;
otherwise it emits the primary's verdict. It provides per-exemplar attribution
for the high-confidence majority and LLM rationale for the low-confidence
minority, while bounding judge invocations. A primary whose scores cluster at
the band mid-point defers everything (vacuous cascade); one whose scores spread
across $[0,1]$ supports meaningful deferral.

\section{\textsc{Principle-Bench} Construction}\label{app:bench}

\paragraph{Generation.}
Each scenario is anchored on one of 22 author-prepared summaries informed by
public FCA material (Finalised Guidance, Dear-CEO letters on cryptoasset
financial promotions, FOS Ombudsman decisions, and FCA-published exemplars; see
the reading list in the supplementary release). \texttt{gpt-oss:120b} is
prompted with the seed, a target verdict, and (channel, product, audience)
constraints, and is forbidden from using any exemplar phrase verbatim. The
target verdict is the ground truth. Paraphrases are LLM rewrites with $\le5$
shared content tokens with any exemplar (stop-words removed; script-verified).
For the adversarial split we attempted 30 keyword-stuffed variants per
polarity-flip target; 28 were retained after the generator's safety alignment
refused 2 prompts framed as ``stuff misleading promotions with compliance
language''; itself a small finding about open-weight safety filters in
compliance-testing contexts. Adversarial variants insert $\ge3$ opposite-polarity
exemplar phrases while preserving the substantive verdict.

\paragraph{Pre-registered rubric.}
Before any scenario was authored or labelled, we committed and tagged two
assessment rubrics (one per principle) operationalising the regulation handbooks
into per-cluster 1--5 Likert criteria and an escalation rule. The rubric commit
and tag \texttt{v0-rubric-prereg} pre-date every label commit in the repository,
a tamper-evident record that the criteria were fixed before scoring.

\paragraph{Limitations.}
Labels are author-assigned via the rubric (no external annotator). We mitigate by
(a)~pre-registering and tagging the rubric before scoring, and (b)~anchoring each
scenario on a public FCA-material summary so the verdict is constrained by
external regulatory practice. We disclose that no external annotator validation
was performed; a regulator panel for inter-annotator agreement is left to
successor work. \emph{Corpus/judge family overlap}: the corpus is authored by
\texttt{gpt-oss:120b} and one judge configuration is the same model; we partly
mitigate with the different-family second judge (\texttt{kimi-k2.6}) and the
inter-judge $\kappa$ analysis (Table~\ref{tab:kappa}). Regenerating the corpus
with a non-overlapping family is left to a successor benchmark.

\section{Full Experimental Setup}\label{app:setup}

\paragraph{Methods.}
\textbf{Keyword}: bag-of-clusters embedder (one dimension per cluster; entries
are L2-normalised pos-minus-neg counts). \textbf{MiniLM}:
\texttt{all-MiniLM-L6-v2} (384-dim). \textbf{BGE}:
\texttt{bge-base-en-v1.5} (768-dim). \textbf{Ollama-mxbai}:
\texttt{mxbai-embed-large} (1024-dim) via local Ollama. \textbf{LLM-judge}:
\texttt{gpt-oss:120b} via Ollama Cloud (temperature~0, seed~42). \textbf{Cascade}:
keyword-primary with LLM-judge fallback when $s\in[0.4,0.6]$. We use the keyword
primary because MiniLM raw scores spanned only $[0.476,0.522]$ on original,
making a 20-sample Platt fit degenerate (slope $a\approx-773$); the keyword
embedder's wider range $[0.39,0.66]$ supports stable band tuning.

\paragraph{Procedure and metrics.}
Calibrators and the cascade band are tuned on a 20-scenario dev subset of
\emph{original}; metrics use the remaining 80 plus all perturbation splits.
For 7--16\% of \emph{original} scenarios the judge returns malformed JSON;
these are excluded from judge cells (the $n$ column). Metrics: accuracy,
precision, recall, F1, per-cluster AUC, false-compliance rate, ECE
\citep{guo2017calibration}, Brier score. All point estimates carry bootstrap
95\% CIs ($10^4$ resamples); paired comparisons use McNemar's exact test.

\section{Full Results}\label{app:results}

\paragraph{Per-cluster AUC (Table~\ref{tab:per_cluster_auc}).}
Keyword and LLM-judge carry the most per-cluster signal; sentence-transformer
embedders are near-random, and several Ollama-mxbai clusters score below 0.5
(0.03--0.09 on three Consumer Duty clusters), i.e.\ an inverted similarity
ordering, evidence that off-the-shelf embeddings used directly can mislead
per-cluster attribution.

\paragraph{Counterfactual faithfulness.}
Tested on all 100 \emph{original} scenarios with keyword and MiniLM backends.
Removing the top-$|\Delta|$ exemplar moves the score by mean $\Delta s=0.018$
(keyword) and $0.0045$ (MiniLM), respectively $10\times$ and $2.3\times$ the
random-exemplar control, and flips the verdict on 4\% [CI 1--8\%] (keyword) and
0\% (MiniLM, reflecting its compressed range). Attribution ranks influence correctly on both;
only the keyword primary produces score movement large enough to flip a binary
verdict on this corpus.

\paragraph{Cascade cost (Table~\ref{tab:cascade}).}
The dev-tuned band $[0.40,0.60]$ defers \cascadeDeferralPct{} of inputs. ECE is
bounded between components and exceeds LLM-judge ECE here; the calibration
cost of mixing poorly calibrated keyword with well-calibrated judge verdicts.
Wall-clock latency is not systematically lower because both shared a
rate-limited endpoint; under independent scheduling expected latency is
$\text{def}\cdot t_\text{judge}+t_\text{primary}\approx0.89\cdot19.7+0.01
\approx17.6$s ($\sim$11\% below the judge). The deployable saving is in the
\emph{count} of judge invocations plus the attribution layer on the
high-confidence majority.

\begin{table}[t]
\centering\small
\begin{tabular}{llccccc}
\toprule
\textbf{Principle} & \textbf{Cluster} & \textbf{Kw} & \textbf{MiniLM} & \textbf{BGE} & \textbf{mxbai} & \textbf{LLM} \\
\midrule
\multirow{4}{*}{CD}
 & Appropriateness   & 0.60 & 0.59 & 0.55 & 0.09 & 0.95 \\
 & Fee fairness      & 0.78 & 0.41 & 0.18 & 0.03 & 0.93 \\
 & Loss capacity     & 0.58 & 0.68 & 0.64 & 0.60 & 0.92 \\
 & Ongoing duty      & 0.68 & 0.45 & 0.53 & 0.37 & 0.92 \\
\midrule
\multirow{4}{*}{COBS}
 & Fee clarity       & 0.72 & 0.58 & 0.20 & 0.06 & 0.96 \\
 & Lang.\ access.    & 0.36 & 0.49 & 0.31 & 0.25 & 0.67 \\
 & Present.\ fair.   & 0.77 & 0.32 & 0.26 & 0.35 & 0.98 \\
 & Risk balance      & 0.54 & 0.33 & 0.49 & 0.34 & 0.99 \\
\bottomrule
\end{tabular}
\caption{Per-cluster AUC on \emph{original}: each cluster's confidence $a_i$ as a
standalone classifier of expected-escalate vs.\ expected-comply.}
\label{tab:per_cluster_auc}
\end{table}

\begin{table}[t]
\centering\small
\begin{tabular}{lcccc}
\toprule
\textbf{Method} & \textbf{Accuracy} & \textbf{ECE} & \textbf{Deferral} & \textbf{Latency (s)} \\
\midrule
Keyword   & 0.53 & 0.152 & n/a & 0.007 \\
LLM-judge & 0.85 & 0.068 & n/a & 19.743 \\
Cascade   & 0.80 & 0.099 & 89\% & 25.308 \\
\bottomrule
\end{tabular}
\caption{Cascade vs.\ components on \emph{original} (averaged over both
principles). Latency is same-environment wall-clock under a shared rate-limited
endpoint; see text for the independent-scheduling bound.}
\label{tab:cascade}
\end{table}

\begin{figure*}[t]
\centering
\includegraphics[width=0.24\textwidth]{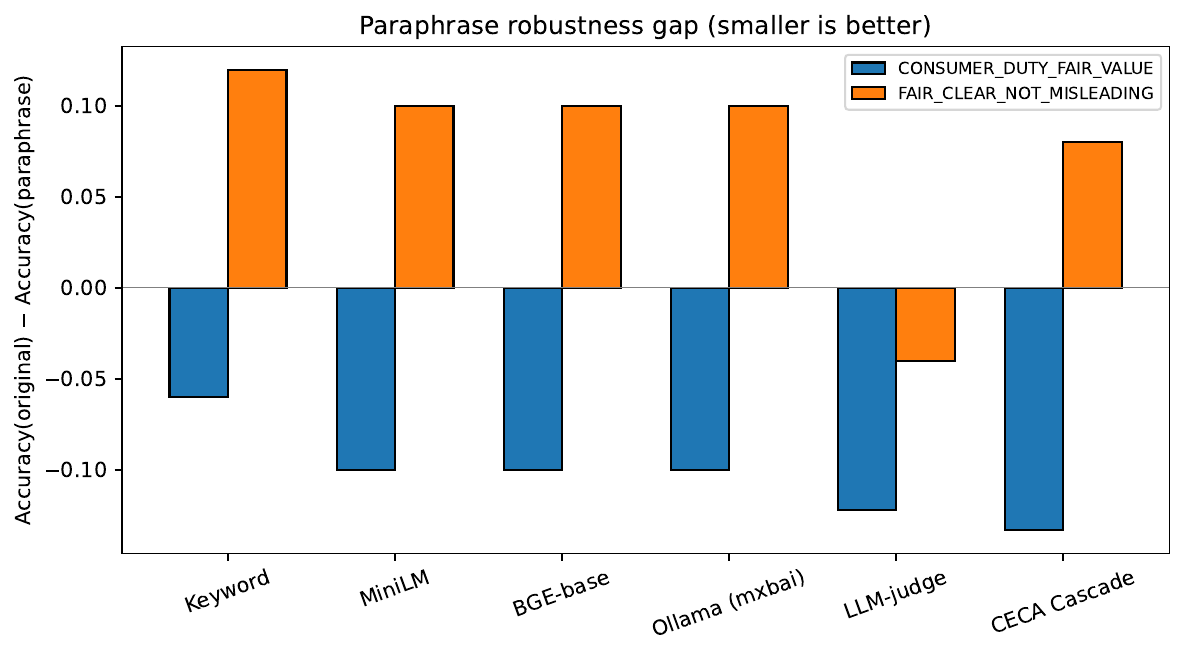}\hfill
\includegraphics[width=0.24\textwidth]{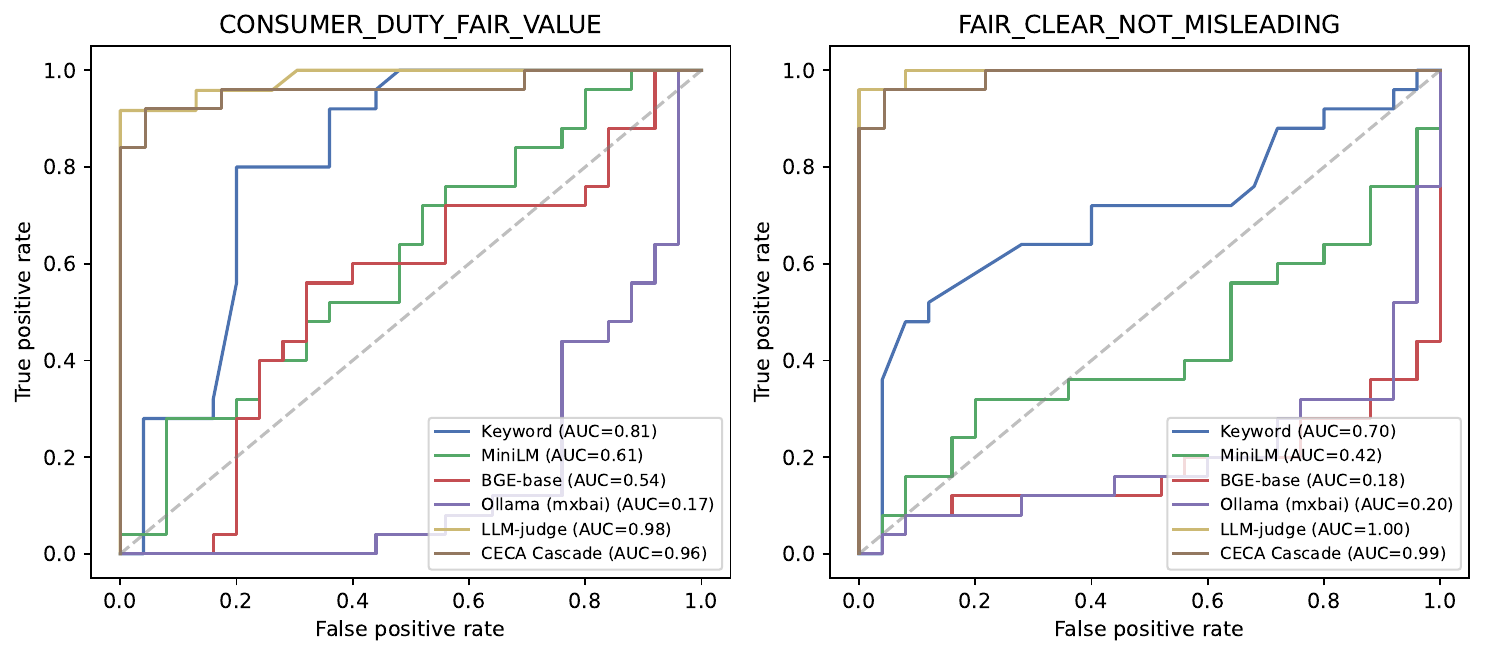}\hfill
\includegraphics[width=0.24\textwidth]{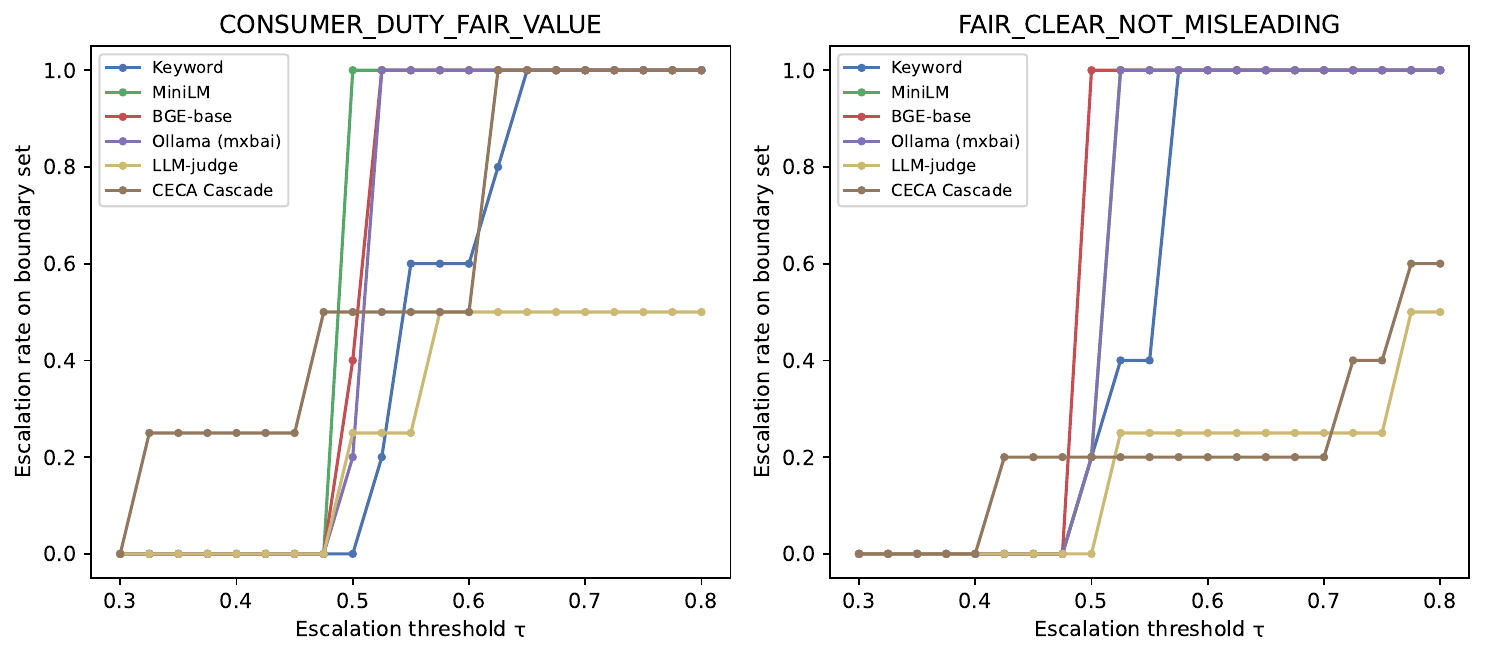}\hfill
\includegraphics[width=0.24\textwidth]{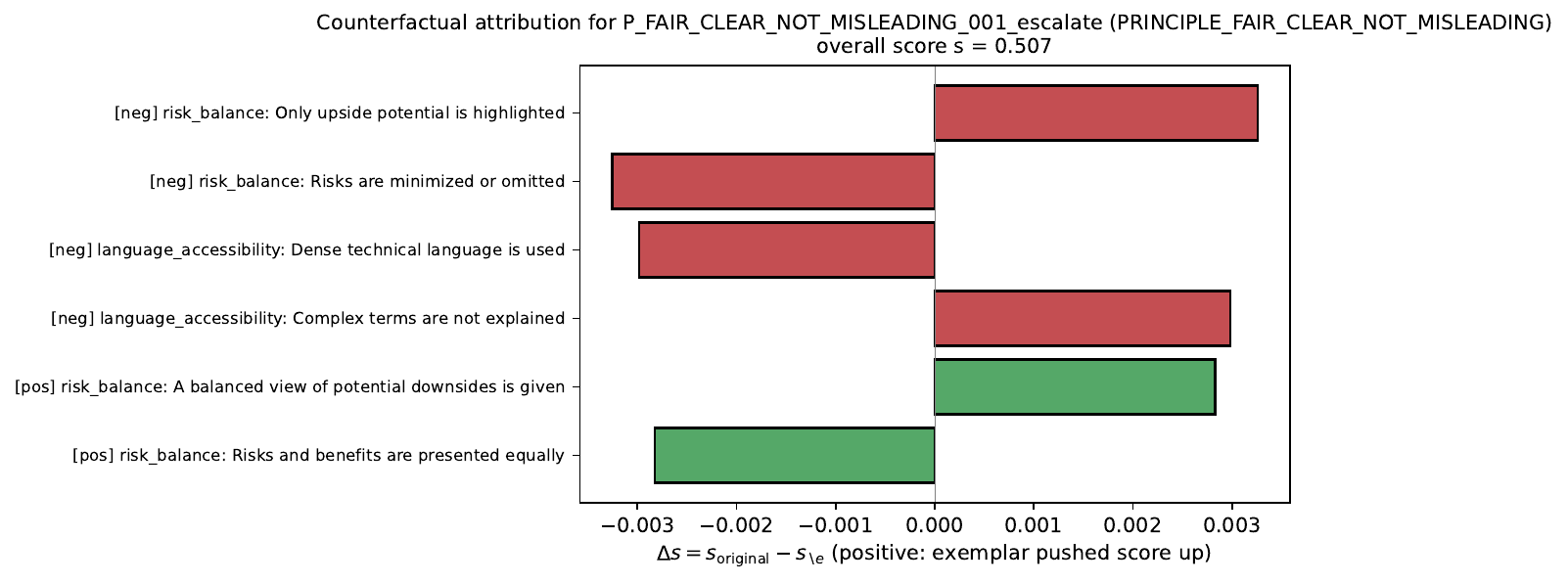}
\caption{Supporting diagnostics (adversarial deception is
Fig.~\ref{fig:adversarial} and reliability diagrams Fig.~\ref{fig:reliability}
in the main text). From left to right: paraphrase robustness gap (smaller is
better); per-method ROC; escalation rate vs.\ $\tau$ over the 10 boundary
scenarios (all methods cross 50\% in $\tau\in[0.45,0.55]$; $n=10$, a
qualitative check); and a worked \textsc{Ceca} attribution example on a
Consumer-Duty-adversarial scenario (signed $\Delta_e$ for the 8
highest-magnitude exemplars, concentrating on ``language accessibility'').}
\label{fig:appfigs}
\end{figure*}

\section{Ethical Considerations}\label{app:ethics}

\textbf{Author-labelled benchmark.} Labels are author-assigned per a
pre-registered rubric; reproducibility is preserved via the rubric and
corpus-freeze tags, but inter-annotator agreement is not reported (no external
annotator). The path to a regulator-validated successor is mapped above.
\textbf{Dual-use.} A principle-assessment tool can triage compliance for
regulators or be used to optimise promotions against the metric while leaving
practice unchanged; our adversarial experiment quantifies this directly. Any
method with a non-trivial deception rate cannot safely be the sole primary
assessor. We recommend periodic human review and public reporting of
method-specific deception rates. \textbf{Synthetic provenance.} Scenarios are
synthetic but anchored on author-prepared summaries of public FCA material; no
verbatim enforcement-document text is used. \textbf{Bias.} Curated exemplars
embed the curator's normative judgements; the pre-registered rubric, open
exemplar set, and counterfactual interface make these inspectable and
contestable, relocating the curation question from a hidden classifier to a
public artefact rather than eliminating it.

\section{Reproducibility}\label{app:repro}

We release the method code, prompts, raw LLM responses, fitted calibrators, the
full benchmark, and the pre-registration tags (\texttt{v0-rubric-prereg},
\texttt{v1-corpus-frozen}) at \url{https://github.com/sarkar-dipankar/principle-bench},
whose commit ordering establishes that the rubric pre-dates all labelling. The full sweep (6
methods~$\times$~4 splits~$\times\le2$ principles) caches one embedding per
(model, exemplar) pair; bootstrap CIs use $10^4$ resamples.

\end{document}